\documentclass[letterpaper,conference,10pt,onecolumn]{ieeetran}
\IEEEoverridecommandlockouts                              
\usepackage{srcltx}
\usepackage{color}
\usepackage{graphicx}
\usepackage{bm}
\usepackage[cmex10]{amsmath}
\usepackage{algorithmic}
\usepackage{algorithm}
\usepackage{amssymb}
\usepackage{cite}
\usepackage{subfigure}
\usepackage{multirow}
\newtheorem{theorem}{Theorem}
\newtheorem{lemma}{Lemma}
\newtheorem{remark}{Remark}

\newtheorem{assumption}{Assumption}

\newtheorem{pf}{Proof}
\usepackage{amsfonts}

\begin{document}

\title{Robust Reduced-Order Model Stabilization for Partial Differential
Equations Based on Lyapunov Theory and Extremum Seeking with
Application to the 3D Boussinesq Equations}
\author{Mouhacine Benosman, Jeff Borggaard, and Boris Kramer
\thanks{ M. Benosman
(m{\_}benosman@ieee.org) is with Mitsubishi Electric Research
Laboratories (MERL), Cambridge, MA 02139, USA. Jeff Borggaard is
with the the Interdisciplinary Center for Applied Mathematics,
Virginia Tech, Blacksburg, VA 24061, USA. B. Kramer is with the
Department of Aeronautics and Astronautics, Massachusetts
Institute of Technology, Cambridge, MA, 02139, USA. }}

\maketitle

\begin{abstract}
We present some results on stabilization for reduced-order models
(ROMs) of partial differential equations. The stabilization is
achieved using Lyapunov theory to design a new closure model that
is robust to parametric uncertainties. The free parameters in the
proposed ROM stabilization method are optimized using a model-free
multi-parametric extremum seeking (MES) algorithm. The 3D
Boussinesq equations provide a challenging numerical test-problem
that is used to demonstrate the advantages of the proposed method.
\end{abstract}

\section{Introduction}
\label{intro} A well known problem in model reduction for partial
differential equations (PDEs) is the so-called {\em stable model
reduction} problem.  The goal is to use Galerkin projection onto a
suitable set of modes to reduce PDEs to a small system of ordinary
differential equations (ODEs), while maintaining the main
characteristics of the original model, such as stability and
prediction precision.

In this paper, we focus on reduced order models obtained by the
method of proper orthogonal decomposition (POD) \cite{HLB98},
which has been widely used to obtain surrogate models of tractable
size in fluid flow applications. However, it has been observed,
e.g.,
\cite{couplet05calibratedROM,kalb07intrinsicPODstab,bui07optimizationROM,ilak10MORGinzburgLandau,kalashnikova2014stabilization},
that POD-ROMs can lose stability.  Maintaining stability is
crucial for any ROM to be accurate over long time intervals.

We address the stable model reduction problem by using closure
models, which are additive, viscosity-like terms introduced in the
ROMs to ensure the stability and accuracy of solutions. Through
Lyapunov theory, we propose a new closure model that is robust to
parametric uncertainties in the model. The obtained closure model
has free parameters, which we auto-tune with a model-free MES
algorithm to optimally match predictions of the PDE model. The
idea of using extremum-seeking to auto-tune closure models has
been introduced in \cite{benosetalacc16}, however, the difference
with this work lies in the new formulation of {\em robust} closure
models. Furthermore, contrary to \cite{benosetalacc16} where the
authors considered the simple case of the Burgers' equation, here
we study the 3D Boussinesq equations, which is a more challenging
test-case and is directly applicable to a number of important
control applications~\cite{kim2015dva}.

Our work extends existing results in the field. Stable model
reduction of Navier-Stokes flow models by adding a nonlinear
viscosity term to the reduced-order model is considered
in~\cite{CNTLDBDN13}.
In~\cite{BDN13,B13}, incompressible flows are stabilized by an
iterative search of the projection modes that satisfy a local
Lyapunov stability condition. An optimization-based approach for
the POD modes of linear models, which solely focused on matching
the outputs of the models is derived in
\cite{bui07optimizationROM,kalashnikova2014stabilization}. Kalb
and Deane \cite{kalb07intrinsicPODstab} added error correction
terms to the reduced-order model for improved accuracy and
stabilization. Moreover, the authors in
\cite{couplet05calibratedROM} calibrated the POD model by solving
a quadratic optimization problem based on three different weighted
error norms.
Stable model reduction for the Navier-Stokes and Boussinesq
equations using turbulence closure models was presented in
\cite{wang2012pod,SI13} and \cite{SB14}, respectively. These
closure models modify some stability-enhancing coefficients of the
reduced-order ODE model using either constant additive terms, such
as the constant eddy viscosity model, or time and space varying
terms, such as Smagorinsky models. The amplitudes of the
additional terms are tuned in such a way to accurately stabilize
the reduced-order model.

However, such closure models do not take into account parametric
uncertainties in the model, and their tuning is not always
straightforward. Our work addresses these issues and proposes a
new closure model in Section~\ref{es-rom-stab} that addresses
parametric uncertainties. Furthermore, we achieve optimal
auto-tuning of this closure model using a learning-based approach,
and is demonstrated using the 3D Boussinesq equations in
Section~\ref{Boussinesq}. To set the stage, the following section
establishes our notation.

\section{Basic Notation and Definitions}
\label{prem} For a vector $q\in \mathbb{R}^n$, the transpose is
denoted by $q^{*}$. The Euclidean vector norm for $q\in
\mathbb{R}^n$  is denoted by $\|\cdot\|$ so that
$\|q\|=\sqrt{q^*q}$. The Frobenius norm of a tensor $A\in
\mathbb{R}^{\otimes_i n_i}$, with elements $a_{\bf i}=a_{i_1\cdots
i_k}$, is defined as $\|A\|_{F}\triangleq\sqrt{\sum_{{\bf i}={\bf
1}}^{\bf n}|a_{\bf i}|^{2}}$. The Kronecker delta function is
defined as: $\delta_{ij}=0,\;\text{for}\;i\neq j$ and
$\delta_{ii}=1$. We call a function analytic in a given set, if it
admits a convergent Taylor series approximation in some
neighborhood of every point of the set. Our PDEs (the Boussinesq
equations) are solved on the unit cube $x\in\Omega = (0,1)^3$ and
$t\in (0,t_f)$.  We shall abbreviate the time derivative by
$\dot{f}(t,x) = \frac{\partial}{\partial t} f(t,x)$, and consider
the following Hilbert spaces: $\mathcal{H}=L^{2}(\Omega)$,
$\mathcal{V}=H_{\rm div}^1(\Omega)\subset (\mathcal{H})^3$ for
velocity and $\mathcal{T}=H^1(\Omega)\subset\mathcal{H}$ for
temperature. Thus, $\mathcal{V}$ is the space of divergence-free
vector fields on $\Omega$ with components in $H^1(\Omega)$.
Dirichlet boundary conditions are also considered in $\mathcal{V}$
and $\mathcal{T}$.  We define the inner product $\langle
\cdot,\cdot\rangle_{\mathcal{H}}$ and the associated norm
$\|\cdot\|_{\mathcal{H}}$ on $\mathcal{H}$ as
$\|f\|_{\mathcal{H}}^{2}=\int_\Omega|f(x)|^{2}dx$, and $\langle
f,g\rangle_{\mathcal{H}}=\int_\Omega f(x)g(x)dx$, for
$f,g\in\mathcal{H}$. A function $T(t,x)$ is in
$L^{2}([0,t_f];\mathcal{H})$ if for each $0\leq t\leq t_f$,
$T(t,\cdot)\in\mathcal{H}$, and
$\int_{0}^{t_f}{\|T(t,\cdot)\|_\mathcal{H}^2} dt<\infty$ with
analogous definitions for the vector valued functions in
$(\mathcal{H})^3$.  To generalize the discussion below, we
consider the abstract Hilbert space $\mathcal{Z}$, and later
specialize to $\mathcal{Z}=\mathcal{V}\oplus\mathcal{T}$ when
considering the Boussinesq equations. Finally, in the remainder of
this paper we consider the stability of dynamical systems in the
sense of Lagrange, e.g., \cite{haddad2008}: A system
$\dot{q}=f(t,q)$ is said to be Lagrange stable if for every
initial condition $q_{0}$ associated with the time instant
$t_{0}$, there exists $\epsilon (q_{0})$, such that
$\|q(t)\|<\epsilon,\;\forall t\geq t_{0}\geq 0$.

\section{Lyapunov-based robust stable model reduction of PDEs}\label{es-rom-stab}
\subsection{Reduced Order PDE Approximation}
We consider a stable dynamical system modeled by a nonlinear
partial differential equation of the form
\begin{equation}\label{general_PDE_chap3}
\dot{z}(t) = \mathcal{F}(z(t)), \qquad z(0) \in\mathcal{Z},
\end{equation}
where $\mathcal{Z}$ is an infinite-dimensional Hilbert space.
Solutions to this PDE can be approximated in a finite dimensional
subspace $\mathcal{Z}^n \subset \mathcal{Z}$ through expensive
numerical discretization, which can be impractical for multi-query
settings such as analysis and design, and even more so for
real-time applications such as prediction and control. In many
systems, including fluid flows, solutions of the PDE may be
well-approximated using only a few suitable (optimal) basis
functions~\cite{HLB98}.

This gives rise to reduced-order modeling through Galerkin
projection, which can be broken down into three main steps: One
first discretizes the PDE using a finite, but large, number of
basis functions, such as piecewise quadratic (for finite element
methods), higher order polynomials (spectral methods), or splines.
In this paper we use the well-established finite element method
(FEM), and refer the reader to the large literature, e.g.,
\cite{gunzburger1989fem}, for details. We denote the approximation
of the PDE solution by $z_n(t,\cdot) \in \mathcal{Z}^n$, where
$\mathcal{Z}^n$ is an $n$-dimensional finite element subspace of
$\mathcal{Z}$.
%
%
Secondly, one determines a small set of spatial basis vectors
$\phi_i(\cdot) \in \mathcal{Z}^n$, $i=1,\ldots,r$, $r\ll n$, that
well approximates the discretized PDE solution with respect to a
pre-specified criterion, i.e.
\begin{equation}\label{x_appro1_chap3}
P_n z(t,x) \approx \Phi q(t) = \sum_{i=1}^{r} q_{i}(t) \phi_i(x).
\end{equation}
Here, $P_n$ is the projection of $z(t,\cdot)$ onto
$\mathcal{Z}^n$, and $\Phi$ is a matrix containing the basis
vectors $\phi_i(\cdot)$ as column vectors.
Note that the dimension $n$, coming from the high fidelity
discretization of the PDE described above, is generally very
large, in contrast to the dimension $r$ of the optimal basis set.
Thirdly, a Galerkin projection yields a ROM for the coefficient
functions $q(\cdot)$ of the form
\begin{equation}\label{ROM1_chap3}
\dot{q}(t) =F(q(t)), \qquad q(0) \in \mathbb{R}^r.
\end{equation}
The function $F:\;\mathbb{R}^r \rightarrow \mathbb{R}^r$ is
obtained using the weak form of the original PDE and Galerkin
projection.

The main challenge in this approach lies in the selection of the
`optimal' basis matrix $\Phi$, and the criterion of optimality
used. There are many model reduction methods to find those basis
functions for nonlinear systems. For example, some of the most
used methods are proper orthogonal decomposition (POD)
\cite{KV02}, dynamic mode decomposition (DMD) \cite{KGBBN15}, and
reduced basis methods (RBM) \cite{veroy2005certified}.

\begin{remark}
    We present the idea of closure models in the
    framework of POD. However, the derivation is not limited to a
    particular basis. Indeed, these closure models can be applied
    to ROMs constructed from other basis functions, such as, DMD.  The motivation comes from the fact that any low-dimensional basis necessarily removes the ability to represent the smallest scale structures in the flow and these structures are responsible for energy dissipation.  The missing dissipation often must be accounted for with an additional modeling term to ensure accuracy and stability of the ROM.
\end{remark}

\begin{remark} \label{remark_Bouss_notation}
    For our Boussinesq example, we could maintain one set of coefficients
    for both velocity and temperature~\cite{podvin2012pod}.  This would be
    reasonable for the class of free-convection problems considered here.
    However, to accommodate forced- and mixed-convection problems, we apply
    the POD procedure below for velocity and temperature data separately.
    We continue to use the framework in (\ref{ROM1_chap3}) and
    consider separate basis functions for velocity, $\phi_i=[(\phi_i^{\bf v})^*; 0^*]^*$
    for $i=1,\ldots,r_{\bf v}$
    and temperature, $\phi_{r_{\bf v}+i}=[0^*; (\phi_i^T)^*]^*$ for $i=1,\ldots,r_T$. The
    different groups of coefficient
    functions $\{q_{i}\}_{i=1}^{r_{\bf v}}$ and $\{q_{i}\}_{i=r_{\bf v}+1}^{r_{\bf v}+r_T}$ (with $r=r_{\bf v}+r_T$) are associated with the independent variables ${\bf v}$ and $T$, respectively.
\end{remark}

\subsection{Proper Orthogonal Decomposition for ROMs}
\label{basic_pod_chap3} POD-based models are most known for
retaining a maximal amount of energy in the reduced
model~\cite{KV02,HLB98}. The POD basis is computed from a
collection of $s$ time snapshots
%
\begin{equation}\label{snap_shot_set_chap3}
\mathcal{S}
=\{z_{n}(t_{1},\cdot),...,z_{n}(t_{s},\cdot)\}\subset\mathcal{Z}^n,
\end{equation}
of the dynamical system, usually obtained from a discretized
approximation of the PDE model in $n$ dimensions. The
$\{t_i\}_{i=1}^s$ are time instances at which snapshots are
recorded, and do not have to be uniform. The \textit{correlation
matrix} $K$ is then defined as
\begin{equation}\label{correlation_matrix_pod_chap3}
{K}_{ij}=\frac{1}{s}\langle
z_{n}(t_{i},\cdot),z_{n}(t_{j},\cdot)\rangle_{\mathcal{H}}
,\;i,j=1,...,s.
\end{equation}
The normalized eigenvalues and eigenvectors of $K$ are denoted by
$\lambda_i$ and $v_i$, respectively. Note that the $\lambda_i$ are
also referred to as the \textit{POD eigenvalues}.
The $i$th \textit{POD basis function} is computed as
\begin{equation}\label{pod_basis_chap3}
\phi_{i}(x)=\frac{1}{\sqrt{s}\sqrt{\lambda_{i}}}\sum_{j=1}^{s}[v_{i}]_j
z_n (t_{j},x),\;i=1,...,r,
\end{equation}
where $r \leq \min \{s,n \}$ is the number of retained POD basis
functions and depends upon the application. The POD basis
functions are orthonormal:
\begin{equation}\label{pod_ortho_chap3}
\langle \phi_{i},\phi_{j}\rangle_{\mathcal{H}}
=\int_\Omega\phi_{i}(x)^*\phi_{j}(x)dx=\delta_{ij},
\end{equation}
where $\delta_{ij}$ denotes the Kronecker delta function.

In this new basis, the solution of the PDE
(\ref{general_PDE_chap3}) can then be approximated by
\begin{equation}\label{pod_proj_chap3}
z_n^{pod}(t,\cdot)=\sum_{i=1}^{r}q_{i}(t) \phi_{i}(\cdot) \ \in
\mathcal{Z}^n,
\end{equation}
where $q_{i},\;i=1,...,r$ are the POD projection coefficients.
%
To find the coefficients $q_i(t)$, the (weak form of the) model
(\ref{general_PDE_chap3}) is projected onto the $r$th-order POD
subspace $\mathcal{Z}^r \subseteq \mathcal{Z}^n \subset
\mathcal{Z}$ using a Galerkin projection in $\mathcal{H}$. In
particular, both sides of equation (\ref{general_PDE_chap3}) are
multiplied by the POD basis functions, where $z(t)$ is replaced by
$z_n^{pod}(t) \in \mathcal{Z}^n$, and then both sides are
integrated over $\Omega$. Using the orthonormality of the POD
basis (\ref{pod_ortho_chap3}) leads to an ODE of the form
(\ref{ROM1_chap3}).
A projection of the initial condition for $z(0)$ can be used to
determine $q(0)$. The Galerkin projection preserves the structure
of the nonlinearities of the original PDE.

\subsection{Closure Models for ROM Stabilization}\label{closure_models_Chap3}  We continue to present the problem of stable model reduction in its general form, without specifying a
particular type of PDE. However, we now assume an affine
dependence of the general PDE  (\ref{general_PDE_chap3}) on a
single physical parameter $\mu$,
\begin{equation}\label{general_PDE2_chap3}
\dot{z}(t) = \mathcal{F}(z(t),\mu), \quad z(0)=z_0
\in\mathcal{Z},\qquad \mu\in\mathbb{R},
\end{equation}
as well as
\begin{assumption}\label{pdestab_assumption1_chap3}
The solutions of the original PDE model (\ref{general_PDE2_chap3})
are assumed to be in $L^{2}([0,\infty);\mathcal{Z})$,
$\forall\mu\in\mathbb{R}$.
\end{assumption}

We further assume that the parameter $\mu$ is critical for the
stability and accuracy of the model, i.e., changing the parameter
can either make the model unstable, or lead to inaccurate
predictions. Since we are interested in fluid dynamics problems,
we can consider $\mu$ as a viscosity coefficient. The
corresponding reduced-order POD model takes the form
(\ref{ROM1_chap3}) and (\ref{pod_proj_chap3}):
\begin{equation}\label{ROM3_chap3}
 \dot{q}(t) =F(q(t),\mu).
\end{equation}
The issue with this Galerkin POD-ROM (denoted POD-ROM-G) is that
the norm of $q$, and hence $z_n^{pod}$, might become unbounded at
a finite time, even if the solution of (\ref{general_PDE2_chap3})
is bounded (Lagrange stable).

The main idea behind the closure modeling approach is to replace
the viscosity coefficient $\mu$ in (\ref{ROM3_chap3})  by a
virtual viscosity coefficient $\mu_{cl}$, whose form is chosen to
stabilize the solutions of the POD-ROM (\ref{ROM3_chap3}).
Furthermore, a penalty term $H(\cdot)$ is added to the original
POD-ROM-G, as follows
\begin{equation}\label{ROM41_chap3}
\dot{q}(t) =F(q(t),\mu)+H(q(t)).
\end{equation}
The term $H(\cdot)$ is chosen depending on the structure of
$F(\cdot,\cdot)$ to stabilize the solutions of
(\ref{ROM41_chap3}). For instance, one can use the Cazemier
penalty model described in \cite{SI13}.

\subsection{Main Result 1: Lyapunov-based Closure Model}
Here we introduce the first main result of this paper, namely a
Lyapunov-based closure model that is robust to parametric
uncertainties. We first rewrite the right-hand side of the ROM
model (\ref{ROM3_chap3}) to isolate the linear viscous term as
follows,
\begin{equation}\label{ROM4_chap3}
F(q(t),\mu) = \widetilde{F}(q(t))+\mu\;Dq(t),
\end{equation}
where $D\in\mathbb{R}^{r\times r}$ represents a constant, negative
definite matrix, and the function $\widetilde{F}(\cdot)$
represents the remainder of the ROM model, i.e., the part without
damping.

We now consider the case where $\widetilde{F}(\cdot)$ might be
unknown, but bounded by a known function. This includes the case
of parametric uncertainties in (\ref{general_PDE2_chap3}) that
produce structured uncertainties in (\ref{ROM4_chap3}). To treat
this case, we use Lyapunov theory and propose a nonlinear closure
model that {\em robustly} stabilizes the ROM in the sense of
Lagrange. Assume that $\widetilde{F}(\cdot)$ satisfies
\begin{assumption}[Boundedness of $\widetilde{F}$\label{pdestabgene_assumption1_chap3}]
The norm of the vector field $\widetilde{F}(\cdot)$ is bounded by
a known function of $q$, i.e., $\|\widetilde{F}(q)\|\leq
\widetilde{f}(q)$.
\end{assumption}
\begin{remark}
Assumption \ref{pdestabgene_assumption1_chap3} allows us to
consider a general class of PDEs and their associated ROMs.
Indeed, all we require is that the right-hand side of
(\ref{ROM3_chap3}) can be decomposed as (\ref{ROM4_chap3}), where
a linear damping term can be extracted and the remaining nonlinear
term $\widetilde{F}$ is bounded.  This could allow for more
general parametric dependencies and includes many structured
uncertainties of the ROM, e.g., a bounded parametric uncertainty
can be formulated in this manner.
\end{remark}

We now present our first main result.
\begin{theorem}
\label{pdestabnlclosurebenos_chap3} Consider the PDE
(\ref{general_PDE2_chap3}) under Assumption
\ref{pdestab_assumption1_chap3}, together with its stabilized ROM
model
\begin{equation}\label{ROM5_chap3}
\dot{q}(t)=\widetilde{F}(q(t))+\mu_{cl}\;Dq(t)+H(q(t)),
\end{equation}
where $\widetilde{F}(\cdot)$ satisfies Assumption
\ref{pdestabgene_assumption1_chap3}, $D \in \mathbb{R}^{r \times
r}$ is negative definite, and $\mu_{cl}$ is given by
\begin{equation}\label{PODROMH_CHAP3}
\mu_{cl}=\mu+\mu_{e}.
\end{equation}
Here $\mu$ is the nominal value of the viscosity coefficient in
(\ref{general_PDE2_chap3}), and $\mu_{e}$ is the additional
constant term. Then, the nonlinear closure model
\begin{equation}\label{nlingenterm1_CHAP3}
H(q)=\mu_{nl}\widetilde{f}(q)~{\rm
diag}(d_{11},...,d_{rr})~q,\quad \mu_{nl}>0
\end{equation}
stabilizes the solutions of the ROM to the invariant set
 $$
\begin{array}{c}
 \hspace{-.5in}\mathcal{S}=\{q\in\mathbb{R}^{r}\;s.t.\;\mu_{cl}\frac{\lambda_{\rm max}(D)\|q\|}{\widetilde{f}(q)}+
\mu_{nl}\|q\|\max\{d_{11}, \ldots, d_{rr}\}+1\geq 0\}.
\end{array}
$$
\end{theorem}
\vspace{.1in}
\begin{pf}
First, we prove that the nonlinear closure model
(\ref{nlingenterm1_CHAP3}) stabilizes the ROM (\ref{ROM5_chap3})
to an invariant set. To do so, we use the following energy-like
Lyapunov function
\begin{equation}
V(q)=\frac{1}{2}{q}^{*}{q}.
\end{equation}
We then evaluate the derivative of $V$ along the solutions of
(\ref{ROM5_chap3}), and use (\ref{nlingenterm1_CHAP3}) and
Assumption \ref{pdestabgene_assumption1_chap3} to write
\begin{align}
\nonumber \dot{V}=&\;{q}^{*}(\widetilde{F}(q)+\mu_{cl}\;Dq+
\mu_{nl}\widetilde{f}(q)~{\rm diag}(d_{11},...,d_{rr})~q)\\
\nonumber \leq&
\;\|q\|\widetilde{f}(q)+\mu_{cl}\|q\|^{2}\lambda_{\rm max}(D)+\mu_{nl}\widetilde{f}(q)\|q\|^{2}\max\{d_{11},...,d_{rr}\}\\
\nonumber \leq&
\;\|q\|\widetilde{f}(q)(1+\mu_{cl}\frac{\lambda(D)_{\rm
max}\|q\|}{\widetilde{f}(q)}+\mu_{nl}\max\{d_{11},...,d_{rr}\}\|q\|).
\end{align}
This shows convergence to the invariant set $\mathcal{S}$.
\hfill$\Box$
\end{pf}
Note that $\lambda_{\rm max}(D)$ and $\max\{d_{11},...,d_{rr}\}$
are negative, thus the sizes of $\mu_{cl}$ and $\mu_{nl}$ directly
influence the size of $\mathcal{S}$.  It is also apparent how the
use of the term $H$ offers robustness when the uncertainty in
$F(\cdot,\cdot)$ is difficult to manage.

\subsection{Main Result 2: MES-based Closure Model Auto-tuning}
\label{MES-closure-tuning_chap3} As discussed in the introduction
as well as in \cite{SB14}, tuning the closure model amplitudes is
important to achieve an optimal stabilization of the ROM. In this
study, we use model-free MES optimization algorithms to tune the
coefficients $\mu_{e}$ and $\mu_{nl}$ of the closure models
presented in Section \ref{closure_models_Chap3}. An advantage of
using MES over other optimization approaches is the auto-tuning
capability that such algorithms allow for, as well as their
ability to continually tune the closure model, {\em even during
online operation of the system}. Indeed, we first use MES to tune
the closure model, but the same algorithm can be coupled to the
real system to continually update the closure model coefficients.

Note that MES-based closure model auto-tuning has many advantages.
First of all, the closure models can be valid for longer time
intervals when compared to standard closure models with constant
coefficients that are identified offline over a (fixed) finite
time interval.  Secondly, the optimality of the closure model
ensures that the ROM obtains the most accuracy for a given
low-dimensional basis, leading to the smallest possible ROM for a
given application.

We begin by defining a suitable learning cost function for the MES
algorithm. The goals of the learning (or tuning) are i.) to
enforce Lagrange stability of the ROM model (\ref{ROM3_chap3}) and
ii.) to ensure that the solutions of the ROM (\ref{ROM3_chap3})
are close to those of the approximation $z_n(t,\cdot)$ to the
original PDE (\ref{general_PDE2_chap3}). The latter learning goal
is important for the accuracy of the solution.

We define the learning cost as a positive definite function of the
norm of the error between the approximate solutions of
(\ref{general_PDE2_chap3}) and the ROM (\ref{ROM41_chap3}),
\begin{equation}\label{Q_pde2_chap3}
\begin{aligned}
Q(\widehat{\bm \mu})&=\widetilde{H}(e_{z}(t,\widehat{\bm \mu})),\\
e_{z}(t,\widehat{\bm \mu})&=z_n^{pod}(t,x;\widehat{ \bm
\mu})-z_n(t,x; {\bm \mu}),
\end{aligned}
\end{equation}
where $\widehat{ \bm \mu}= [\widehat\mu_e,\widehat\mu_{nl}]^*
\in\mathbb{R}^2$ denotes the learned parameters, and
$\widetilde{H}(\cdot)$ is a positive definite function of $e_{z}$.
Note that the error $e_{z}$ could be computed offline using
solutions of the ROM (\ref{ROM41_chap3}) and approximate solutions
of the PDE (\ref{general_PDE2_chap3}). The error could be also
computed online where the $z_n^{pod}(t,x;\widehat{\bm \mu})$ is
obtained from solving the model (\ref{ROM41_chap3}) online, but
the $z_n(t,x; {\bm \mu})$ could be replaced by real measurements
of the system at selected spatial locations $\{x_i\}$. The latter
approach would circumvent the FEM model, and directly operate on
the system, making the reduced order model more consistent with
respect to the operating plant.

A practical way to implement the MES-based tuning of $\widehat{\bm
\mu}$, is to begin with an offline tuning of the closure model.
One then uses the obtained ROM (with the optimal values of
$\widehat{\bm \mu}$, namely ${\bm \mu}^{\rm{opt}}$) in the online
operation of the system, e.g., control and estimation. We can then
fine-tune the ROM online by continuously learning the best value
of $\widehat{\bm \mu}$ at any given time during the operation of
the system.

To derive formal convergence results, we introduce some classical
assumptions on the learning cost function.
\begin{assumption} \label{robustmesass1_pdestab_chap3}
The cost function $Q(\cdot)$ in (\ref{Q_pde2_chap3}) has a local
minimum at $\widehat{\bm \mu}= {\bm \mu}^{\rm opt}$.
\end{assumption}
\begin{assumption} \label{robustmesass2_pdestab_chap3}
The cost function $Q(\cdot)$ in (\ref{Q_pde2_chap3}) is analytic
and its variation with respect to ${\bm \mu}$ is bounded in the
neighborhood of ${\bm \mu}^{\rm opt}$, i.e., $\|\nabla_{\bm \mu}
{Q}({\widetilde{\bm \mu}})\|\leq\xi_{2},\;\xi_{2}>0$, for all
$\widetilde{\bm \mu}\in\mathcal{N}({\bm \mu}^{\rm opt})$, where
$\mathcal{N}({\bm \mu}^{\rm opt})$ denotes a compact neighborhood
of ${\bm \mu}^{\rm opt}$.
\end{assumption}

Under these assumptions the following lemma holds.
\begin{lemma}\label{pdestab_lemma1_chap3}
Consider the PDE (\ref{general_PDE2_chap3}) under Assumption
\ref{pdestab_assumption1_chap3}, together with its ROM model
(\ref{ROM5_chap3}), (\ref{PODROMH_CHAP3}), and
(\ref{nlingenterm1_CHAP3}). Furthermore, suppose the closure model
amplitudes $ {\widehat{\bm \mu}} = [\mu_{e},\;\mu_{nl}]^*$ are
tuned using the MES algorithm
\begin{equation}
\begin{aligned}
\dot{y}_{1}(t)  &=a_{1}\sin \left (\omega_{1} t+\frac{\pi}{2} \right )Q( \widehat{\bm \mu}),\\
\widehat{\mu}_{e}(t)&=y_{1}+a_{1}\sin \left (\omega_{1} t-\frac{\pi}{2} \right ),\\
\dot{y}_{2}(t)  &=a_{2}\sin \left (\omega_{2} t+\frac{\pi}{2} \right )Q( \widehat{\bm \mu}),\\
\widehat{\mu}_{nl}(t) &=y_{2}+a_{2}\sin \left (\omega_{2}
t-\frac{\pi}{2} \right ), \label{pdestab_mes_2_chap3}
\end{aligned}
\end{equation}
where $\omega_{\rm max}=\max(\omega_{1},\omega_{2})>\omega^{\rm
opt}$, $\omega^{\rm opt}$ large enough, and $Q(\cdot)$ is given by
(\ref{Q_pde2_chap3}). Let $e_{\bm \mu}(t):=[{{\bm \mu}_{e}}^{\rm
opt}-{\widehat{\bm \mu}}_{e}(t),{{\bm \mu}_{nl}}^{\rm
opt}-{\widehat{\bm \mu}}_{nl}(t)]^*$ be the error between the
current tuned values, and the optimal values ${\bm \mu}_{e}^{\rm
opt},\; {\bm \mu}_{nl}^{\rm opt}$. Then, under Assumptions
\ref{robustmesass1_pdestab_chap3}, and
\ref{robustmesass2_pdestab_chap3}, the norm of the distance to the
optimal values admits the following bound
\begin{equation}
\begin{array}{l}
\|e_{\bm \mu}(t)\|\leq\frac{\xi_{1}}{\omega_{\rm
max}}+\sqrt{a_{1}^{2}+a_{2}^{2}},\;t\rightarrow\infty,
\end{array}\label{pdestab_mes_bound3_chap3}
\end{equation}
where $a_{1},\;a_{2}>0,\;\xi_{1}>0$, and the learning cost
function approaches its optimal value within the following
upper-bound
\begin{equation}
\begin{array}{l}
\|Q({\widehat{\bm \mu}})-Q({\bm \mu}^{\rm
opt})\|\leq\xi_{2}(\frac{\xi_{1}}{\omega}+\sqrt{a_{1}^{2}+a_{2}^{2}}),
\end{array}\label{pdestab_mes_bound4_chap3}
\end{equation}
as $t\rightarrow\infty$, where $\xi_{2}=\max_{ {\bm \mu} \in
\mathcal{N}({\bm \mu}^{\rm opt})}\|\nabla_{\bm \mu}{Q}({\bm
\mu})\|$.
\end{lemma}
\begin{pf}
Based on Assumptions \ref{robustmesass1_pdestab_chap3}, and
\ref{robustmesass2_pdestab_chap3}, the extremum seeking nonlinear
dynamics (\ref{pdestab_mes_2_chap3}), can be approximated by a
linearly averaged dynamic model (using an averaging approximation
over time, \cite{R00}, p. 435, Definition 1). Furthermore,
$\exists\; \xi_{1},\;\omega^{\rm opt}$, such that for all
$\omega>\omega^{\rm opt}$, the solution of the averaged model
${\widehat {\bm \mu}}_{\rm aver}(t)$ is locally close to the
solution of the original MES dynamics, and satisfies (\cite{R00},
p. 436 )
$$
\begin{array}{c}
\|\widehat {\bm \mu}(t)-{\bf d}(t)-{\widehat {\bm \mu}}_{\rm
aver}(t)\| \leq\frac{\xi_{1}}{\omega},\;\xi_{1}>0,\;\forall t\geq
0,
\end{array}
$$
with $ {\bf d}(t)=[{\rm a_1\;sin}(\omega_1 t-\frac{\pi}{2}),{\rm
a_2\;sin}(\omega_2 t-\frac{\pi}{2})]^*$. Moreover, since
$Q(\cdot)$ is analytic it can be approximated locally in
$\mathcal{N}({\bm \mu}^{\rm opt})$ with a quadratic function,
e.g., Taylor series up to second order, which leads to
(\cite{R00}, p. 437 )
$$
\lim_{t\rightarrow\infty}{\widehat {\bm \mu}}_{\rm aver}(t)= {\bm
\mu}^{\rm opt}.
$$
Based on the above, we can write
$$
\|\widehat {\bm \mu}(t)- {\bm \mu}^{\rm opt}\|-\| {\bf d}(t)\|\leq
\|\widehat {\bm \mu}(t)- {\bm \mu}^{\rm opt}- {\bf d}(t)\|
\leq\frac{\xi_{1}}{\omega},\\
$$
so that
$$
\| \widehat {\bm \mu}(t)- {\bm \mu}^{\rm opt}\|
\leq\frac{\xi_{1}}{\omega}+\| {\bf d}(t)\| \quad,
t\rightarrow\infty,
$$
which implies
$$
\|\widehat {\bm \mu}(t) - {\bm \mu}^{\rm opt}\|\leq
\frac{\xi_{1}}{\omega}+\sqrt{a_{1}^{2}+a_{2}^{2}},\;\xi_{1}>0,\;t\rightarrow\infty.
$$
Next, the cost function upper-bound is easily obtained from the
previous bound, using the fact that $Q(\cdot)$ is locally
Lipschitz, with Lipschitz constant $\xi_{2}=\max_{ {\bm
\mu}\in\mathcal{N}({\bm \mu}^{\rm opt})}\|\nabla_{\bm \mu} Q({\bm
\mu})\|$.
 \hspace{+0cm}$\Box$
\end{pf}

\section{The 3D Boussinesq equation}
\label{Boussinesq} As an example application of our approach, we
consider the 3D incompressible Boussinesq equations that describe
the evolution of velocity ${\bf v}$, pressure $p$, and temperature
$T$ of a fluid. This system serves as a model for the flow of air
in a room. The coupled equations reflect the conservation of
momentum, mass, and energy, respectively
\begin{align}\label{Boussinesq1}
  \rho \left(\frac{\partial {\bf v}}{\partial t} + {\bf v}\cdot\nabla{\bf v} \right) &=
  -\nabla p + \nabla\cdot {\mathbf \tau}({\bf v})
  +\rho {\bf g},\\
  \nabla \cdot {\bf v} &= 0,\\
  \rho c_p \left(\frac{\partial T}{\partial t} + {\bf v}\cdot\nabla T\right) &=
  \nabla \left( \kappa \nabla T \right),
\end{align}
where the buoyancy force is driven by changes in density $\rho =
\rho_0 + \Delta \rho$, and is modeled as perturbations from the
nominal temperature $T_0$ using the perfect gas law $\Delta\rho
{\bf g} \approx -\rho_0 \beta \left( T-T_0 \right) {\bf g}$,
$\beta=1/T_0$, and the term $\rho_0 {\bf g}$ is absorbed into the
pressure.  The viscous stress is $\tau({\bf v}) = \rho\nu \left(
\nabla{\bf v} + \nabla{\bf v}^T\right)$ with kinematic viscosity
$\nu$ and thermal conductivity $\kappa$, and the gravitational
acceleration is ${\bf g} = -g\widehat{\bf e}_3$. One typically
non-dimensionalizes these equations depending on the application
at hand.  For this study, we perform non-dimensionalization as
follows. By introducing a characteristic length $L$,
characteristic velocity ${\bf v}_0$, wall temperature $T_w$, and
defining $\widetilde{x} = \frac{x}{L}$, $\widetilde{t} = \frac{t
{\bf v}_0}{L}$, $\widetilde{\bf v} = \frac{\bf v}{ {\bf v}_0}$,
$\widetilde{p} = \frac{p}{\rho {\bf v}_0^2}$, and $\widetilde{T} =
\frac{T-T_0}{T_w-T_0}$ we can reduce the number of free parameters
to three.  These are the Reynolds number ${\rm Re}
= \frac{ {\bf v}_0 L }{\nu}$, the Grashof number ${\rm Gr} =
\frac{ g \beta (T_w-T_0) L^3 }{ \nu^2 }$, and the Prandtl number
${\rm Pr} = \frac{\nu}{k/\rho c_p}$.  Thus,
\begin{align}\label{Boussinesq2}
  \frac{\partial {\bf v}}{\partial t} + {\bf v}\cdot\nabla {\bf v} &=
  -\nabla p + \nabla \cdot \tau({\bf v})
  + \frac{{\rm Gr}}{{\rm Re}^2} T\;\widehat{\bf e}_3, \\
  \nabla \cdot {\bf v} &= 0, \\
\label{Boussinesq_en}
  \frac{\partial T}{\partial t} + {\bf v}\cdot \nabla T &= \nabla\cdot\left(\frac{1}{\rm Re Pr}
  \nabla T \right),
\end{align}
where $\tau({\bf v}) = \frac{1}{\rm Re}(\nabla{\bf v}+\nabla{\bf
v}^T)$ and we have dropped the tilde notation. 
%

Following a Galerkin projection onto the subspace spanned by the
POD basis functions, the Boussinesq equation is reduced to a POD
ROM with the following structure, e.g., \cite{CNTLDBDN13}

\begin{align} \label{NS-PODROM1-CHAP3}
\dot{q}(t) & = \mu\;D\;q(t)+[Cq(t)]{q(t)},\\
{\bf v}(x,t) & ={\bf v}_{0}(x)+\sum_{i=1}^{r_{\bf v}} q_i(t) \phi_i^{\bf v}(x),\\
T(x,t) & =T_{0}(x)+\sum_{i=r_{\bf v}+1}^{r_T+r_{\bf v}} q_{i}(t)
\phi_i^T(x),
\end{align}
where $\mu>0$ is the viscosity, i.e., the inverse of the Reynolds
number, $D$ is a negative definite matrix with diagonal blocks
corresponding to the viscous stress and thermal diffusion (scaled
by {\rm Pr} to extract the parameter $\mu$) and $C$ is a
three-dimensional tensor corresponding to the convection terms in
(\ref{Boussinesq2}) and (\ref{Boussinesq_en}). Recall the
notational setting Remark \ref{remark_Bouss_notation}, where we
formulated the Boussineq equations in the general framework of
(\ref{ROM1_chap3}).
We notice that this POD-ROM has mainly a linear term and two
quadratic terms, so that it can be written in the form
(\ref{ROM4_chap3}), with
$$\widetilde{F}(q)=[Cq]{q}.$$
If we consider bounded parametric uncertainties for the entries of
$C$, we can write
$$\widetilde{F}(q)=[(C+\Delta C)q]{q},$$
where $\|C+\Delta C\|_{F}\leq c_{\rm max}$, we have the
upper-bound
$$\|\widetilde{F}(q)\|\leq \widetilde{f}(q)\equiv c_{\rm max}\|q\|^{2}.$$
In this case the nonlinear closure model
(\ref{nlingenterm1_CHAP3}) is
\begin{equation}\label{nlingenterm2_CHAP3}
H(q)=\mu_{nl}c_{\rm max} \|q\|^{2} {\rm diag}(d_{11},...,d_{rr})q,
\end{equation}
for $\mu_{nl}>0$ with $d_{ii},\;i=1,...,r$ being the diagonal
elements of $D$.

\subsection{Boussinesq equation
MES-based POD ROM stabilization}  We consider the
Rayleigh-B\'enard differential-heated cavity problem, modeled with
the 3D Boussinesq equations
(\ref{Boussinesq2})--(\ref{Boussinesq_en}) with the following
parameters and boundary conditions.  The unit cube was discretized
with 495k quadratic tetrahedral elements with 611k nodes leading
to 1.83M velocity degrees of freedom and 611k temperature degrees
of freedom.  Thus, $n\approx 2.4\times 10^6$.  The velocity was
taken as zero on the boundary and the temperature was specified at
$\pm 0.5$ on the $x$-faces and taken as homogeneous Neumann on the
remaining faces.  The non-dimensional parameters were taken as
${\rm Re}=4.964\times 10^4$, ${\rm Pr}=0.712$, and ${\rm
Gr}=7.369\times 10^7$, reasonable values in a quiet room.  The
simulation was run from zero velocity and temperature and
snapshots were collected to $t_f=78$ seconds.

We apply the results of Theorem \ref{pdestabnlclosurebenos_chap3}
and Lemma \ref{pdestab_lemma1_chap3} to this problem. In this case
we use $8$ POD basis functions for each variable, for the POD
model (POD-ROM-G). The upper bounds on the uncertainties in the
matrix and tensor entries are assumed to be $c_{\rm max}=10$. The
two closure model amplitudes ${\widehat{\bm \mu} } = [\mu_{e},
\mu_{nl}]^*$ are tuned using the discrete version of the MES
algorithm (\ref{pdestab_mes_2_chap3}), given by
\begin{equation}
\begin{aligned}
y_{1}(k+1) & = y_{1}(k)+a_{1}\Delta t\sin \left (\omega_{1} k\Delta t+\frac{\pi}{2}\right ) Q(\widehat{\bm \mu}  ),\\
\widehat{\mu}_{e}(k+1) & = y_{1}(k+1)+a_{1}\;\sin \left (\omega_{1} k\Delta t-\frac{\pi}{2} \right ),\\
y_{2}(k+1) & = y_{2}(k)+a_{2}\Delta t\sin \left (\omega_{2} k\Delta t+\frac{\pi}{2}\right ) Q(\widehat{\bm \mu}),\\
\widehat{\mu}_{nl}(k+1) & =y_{2}(k+1)+a_{2}\;\sin \left
(\omega_{2} k\Delta t-\frac{\pi}{2} \right ),
\label{pdestab_mes_2discrt_chap3}
\end{aligned}
\end{equation}
where $y_{1}(0)=y_{2}(0)=0$, $k=0,1,2,...$ is the number of
learning iterations, and $\Delta t$ is the time increment. We use
MES parameter values: $a_{1}=0.08\;[-],\;\omega_{1}=10\;
[\frac{\rm rad}{ \rm
sec}],\;a_{2}=10^{-7}\;[-],\;\omega_{2}=50\;[\frac{\rm rad}{ \rm
sec}]$. The learning cost function is chosen as
\begin{equation}\label{Re_estim_Q1_chap3_}
Q({\bm \mu }) = \int_{0}^{t_{f}} \langle e_{T},e_{T} \rangle_{\cal
H} dt  +  \int_{0}^{t_{f}} \langle e_{\bf v},e_{\bf v} \rangle_{
({\cal H})^3 } dt.
\end{equation}
Moreover, $e_{T}=P_rT_{n}-T^{pod}_{n},\;e_{\bf v}=P_r{\bf
v}_{n}-{\bf v}^{pod}_{n}$ define the errors between the projection
of the true model solution onto the POD space ${\mathcal Z}^r$ and
the POD-ROM solution
for temperature and velocity, respectively. 

\begin{figure}
 \center
\includegraphics[width=0.5\linewidth]{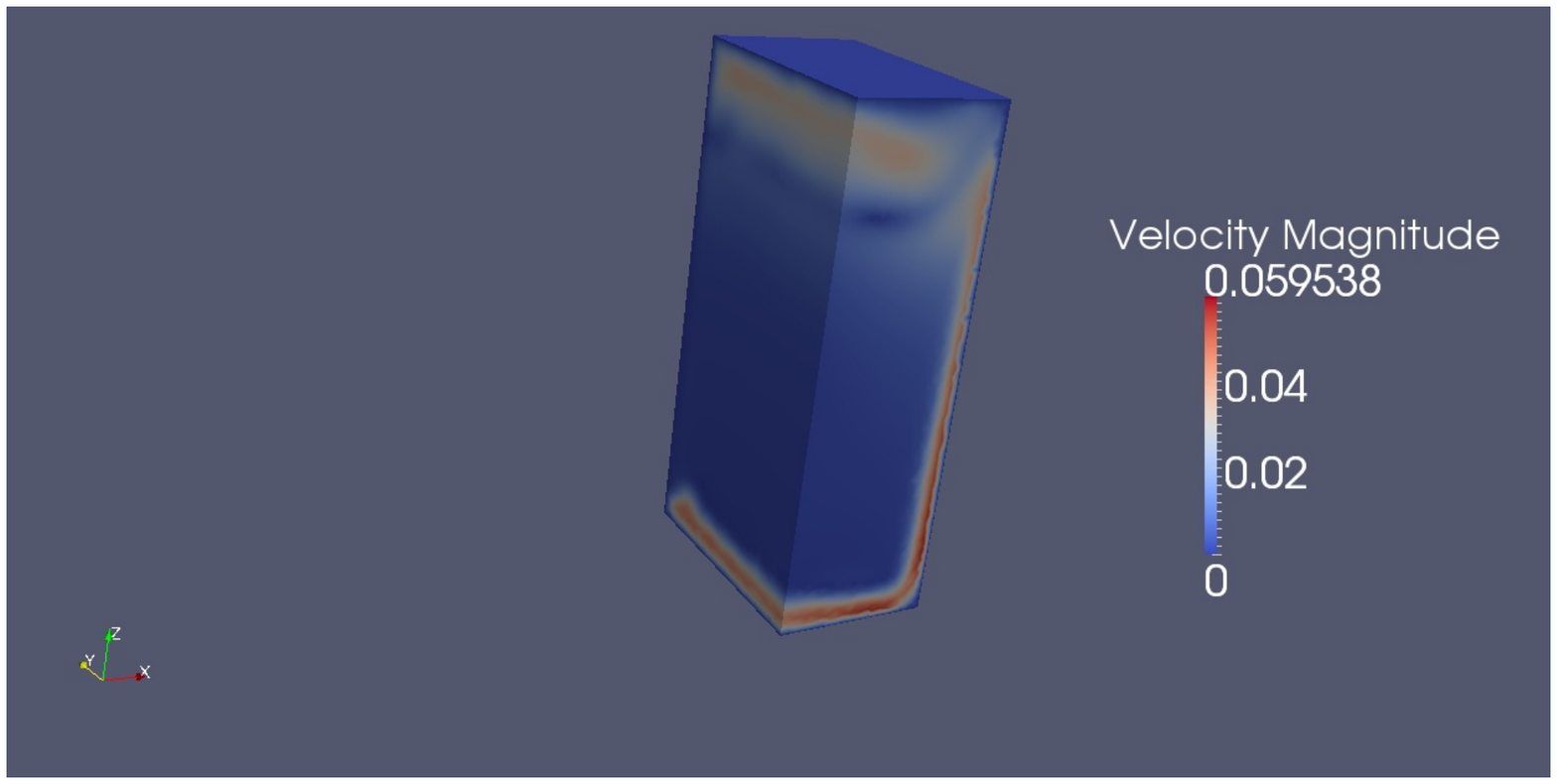}
  \vspace{-8cm}\caption{True velocity profile}
  \label{TrueV}
\end{figure}

\begin{figure}
 \center
\includegraphics[width=0.5\linewidth]{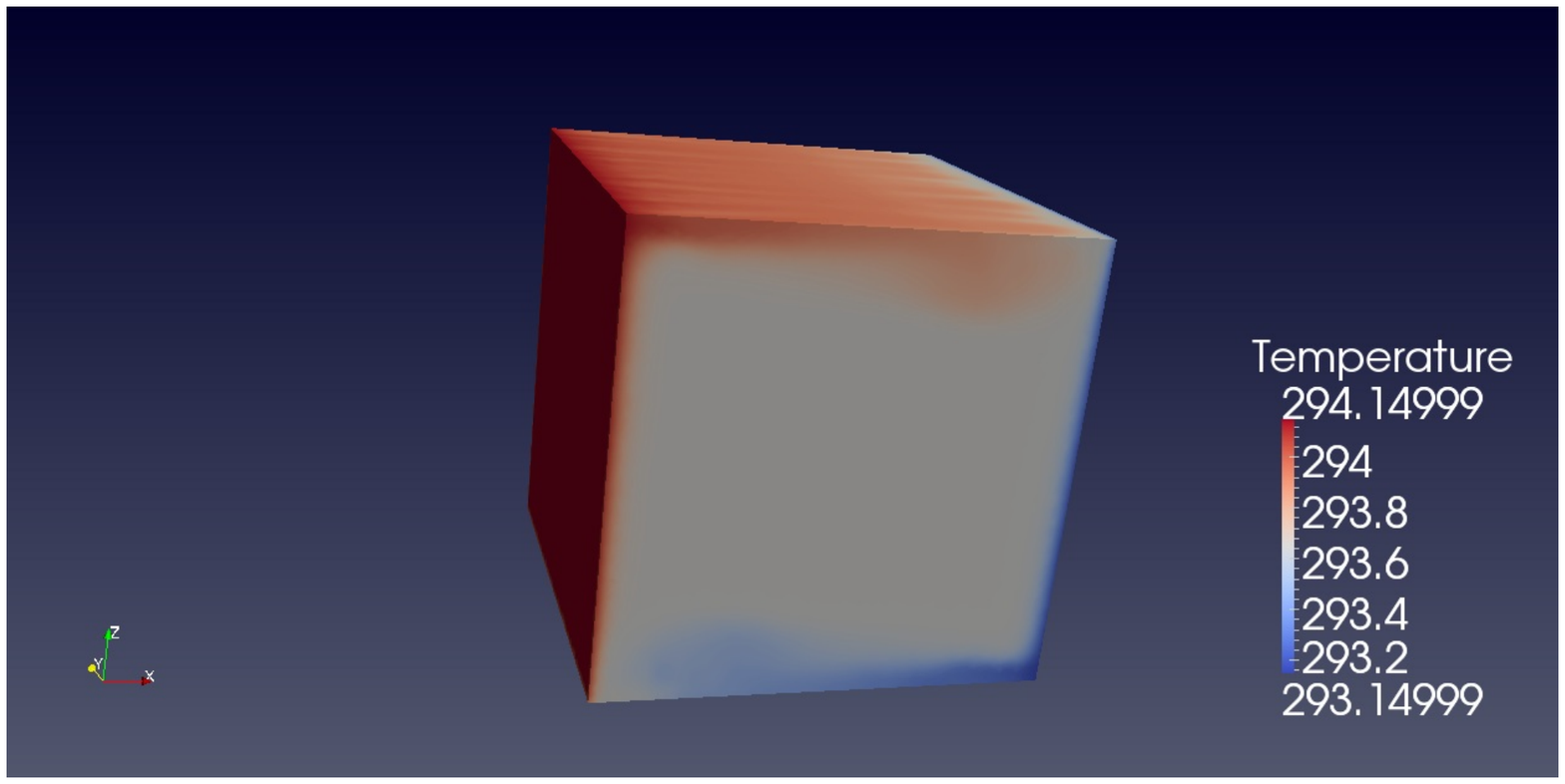}
  \vspace{-8cm}\caption{True temperature profile}
  \label{TrueT}
\end{figure}

\begin{figure}
 \center
\includegraphics[width=0.5\linewidth]{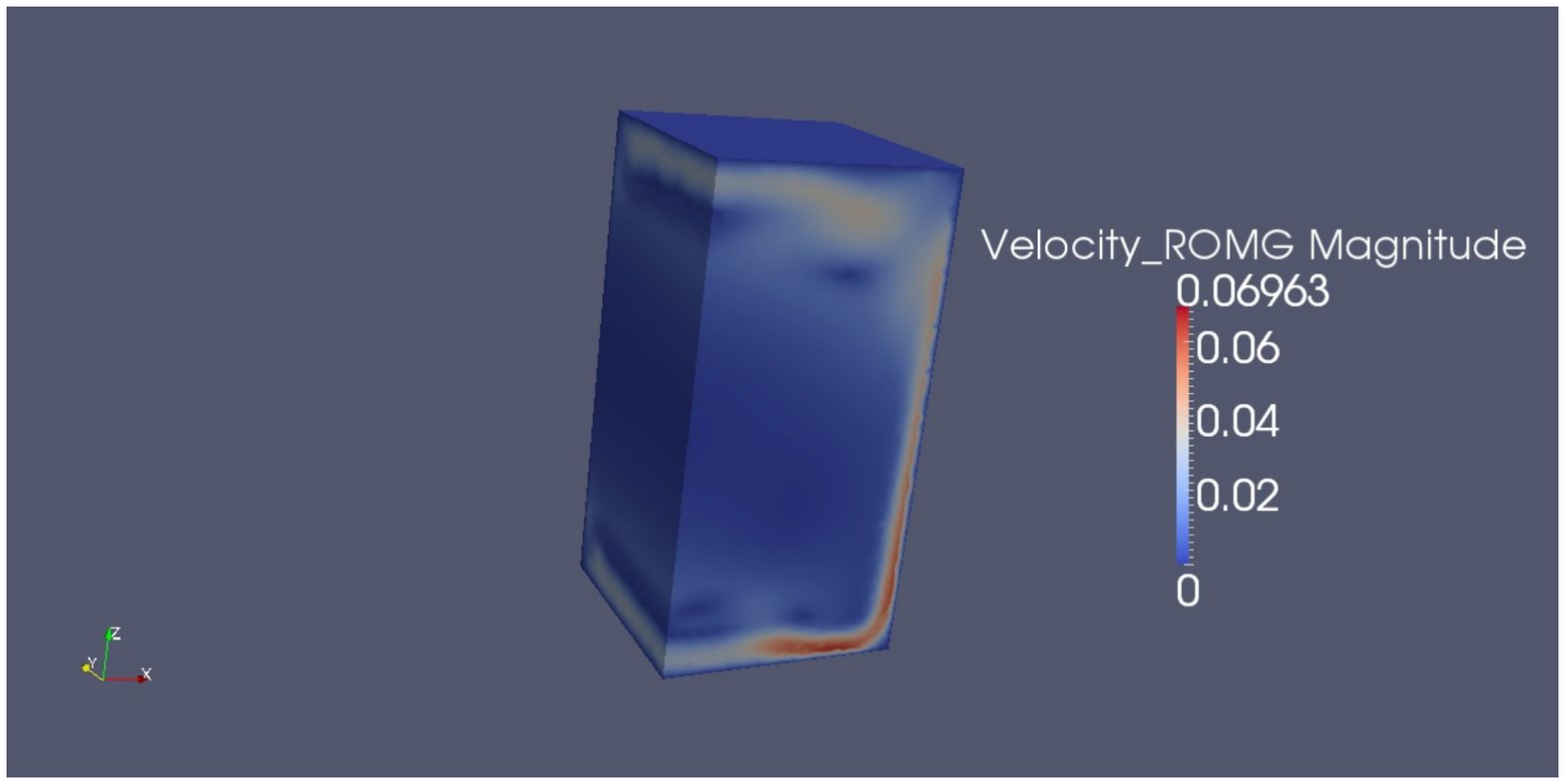}
  \vspace{-8cm}\caption{ROM-G velocity profile}
  \label{PODGV}
\end{figure}

\begin{figure}
 \center
\includegraphics[width=0.5\linewidth]{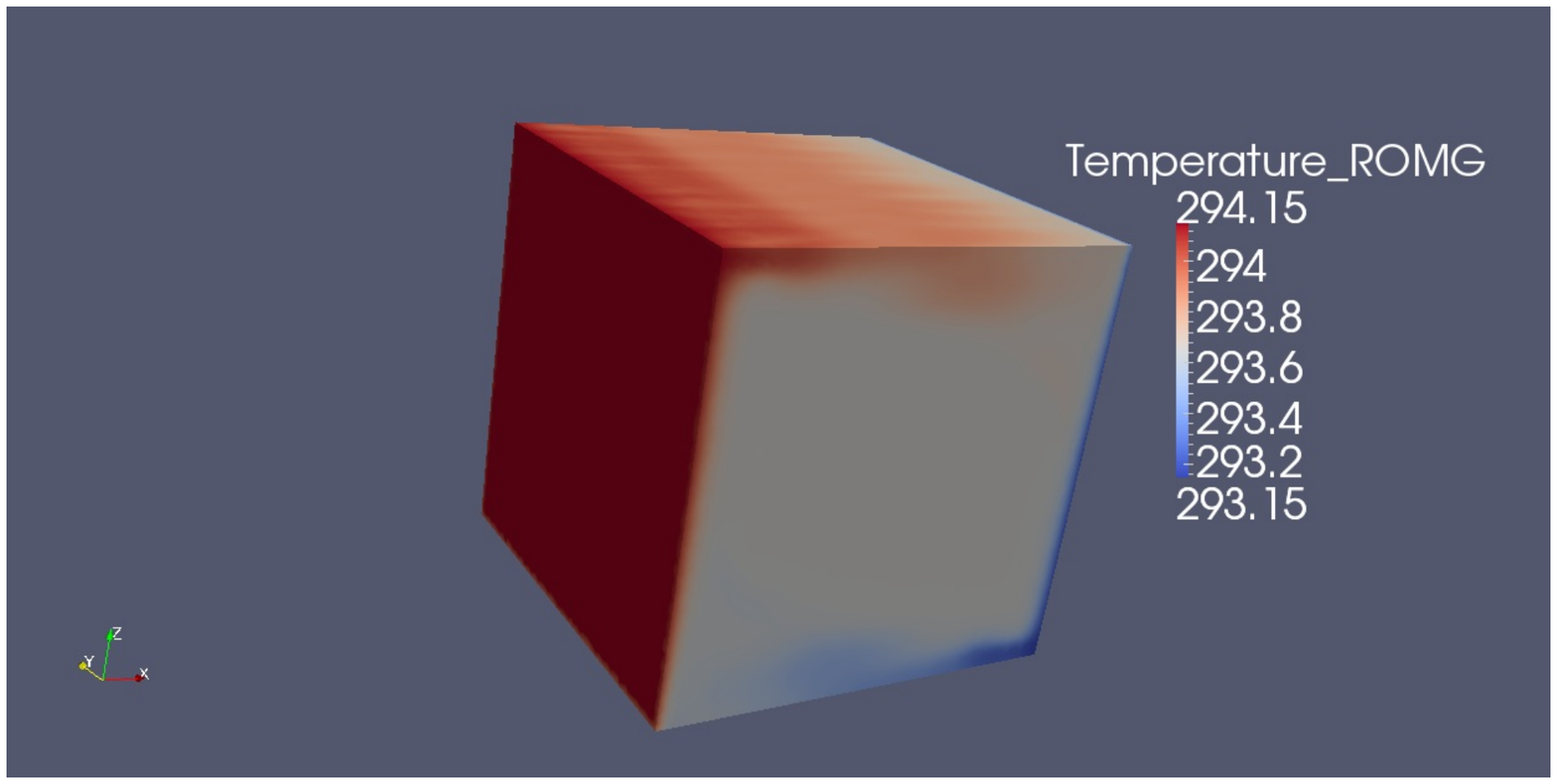}
  \vspace{-8cm}\caption{ROM-G temperature profile}
  \label{PODGT}
\end{figure}

\begin{figure}
 \center
\includegraphics[width=0.5\linewidth]{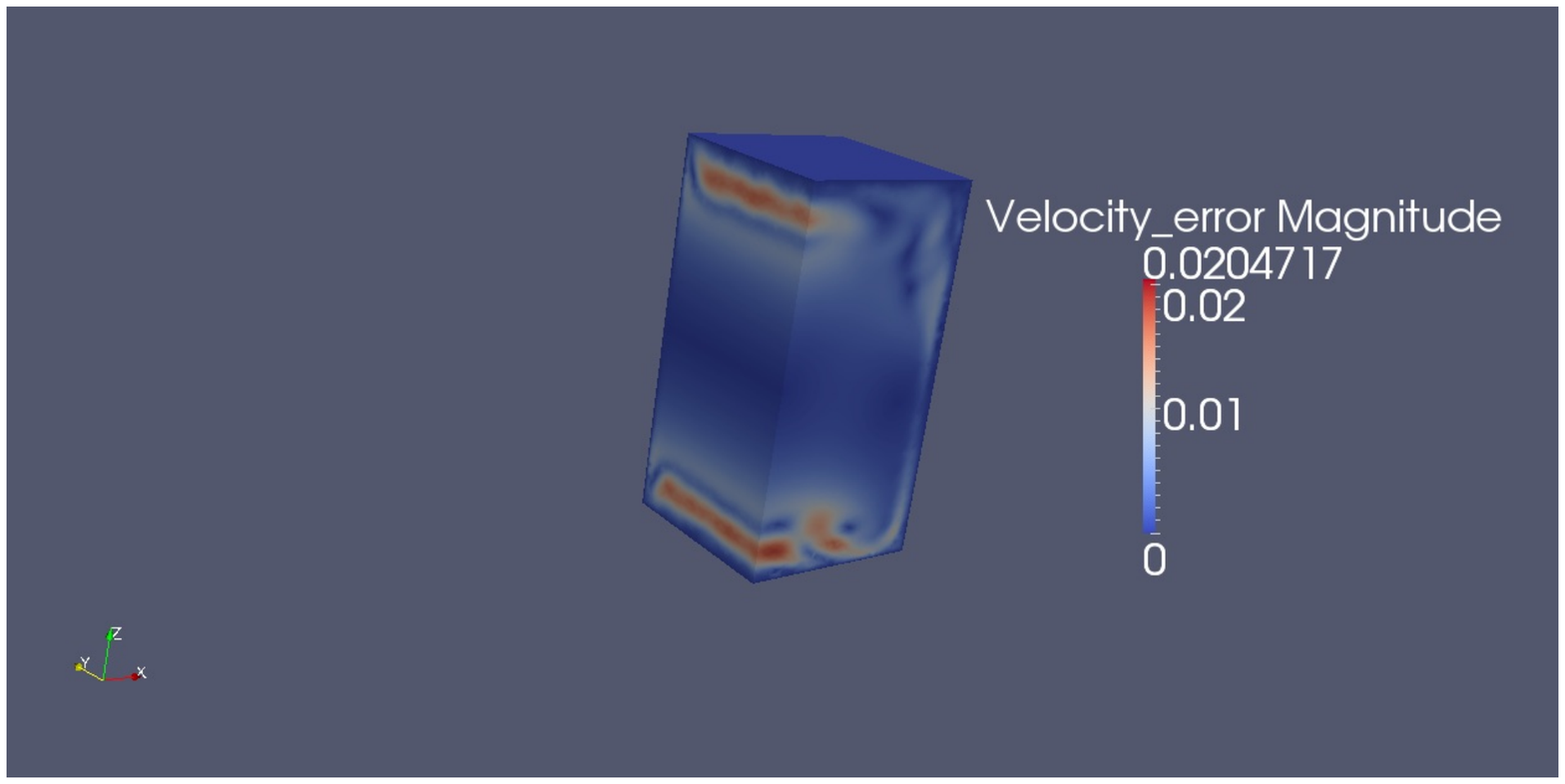}
  \vspace{-8cm}\caption{ROM-G velocity error profile}
  \label{errorVG}
\end{figure}

\begin{figure}\center
\includegraphics[width=0.4\linewidth]{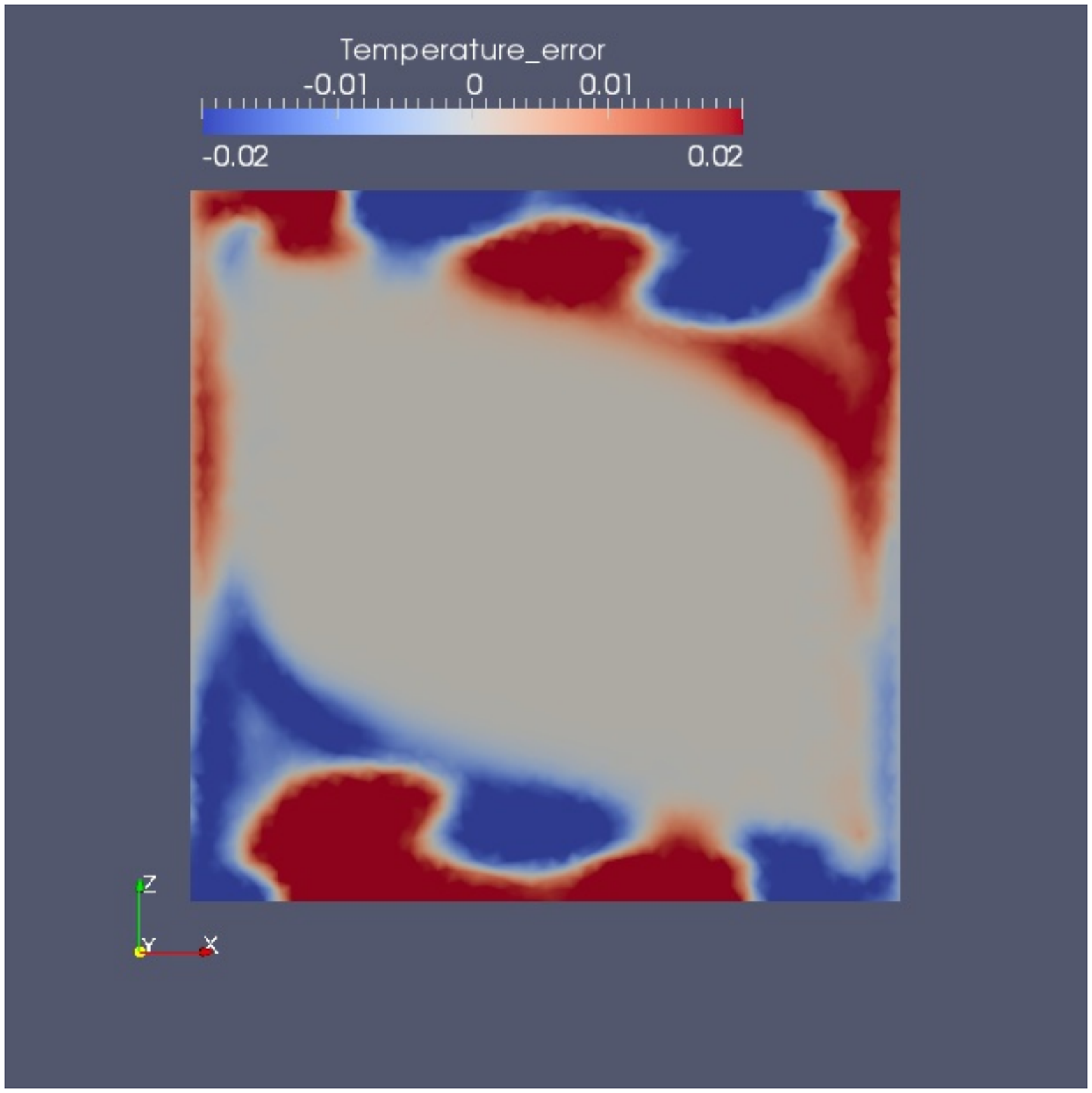}
  \vspace{-3cm}\caption{ROM-G temperature error profile}
  \label{errorTG}
\end{figure}

We first report in Figures \ref{TrueV}, \ref{TrueT} the true
velocity and temperature solutions. Figures \ref{PODGV},
\ref{PODGT} show the solutions obtained at $t=50\sec$ with the
nominal Galerkin ROM, with no closure model. We then report the
errors between the true solutions and the POD-ROM-G solutions in
Figures \ref{errorVG}, and \ref{errorTG}.

\begin{figure}
 \center
\includegraphics[width=0.5\linewidth]{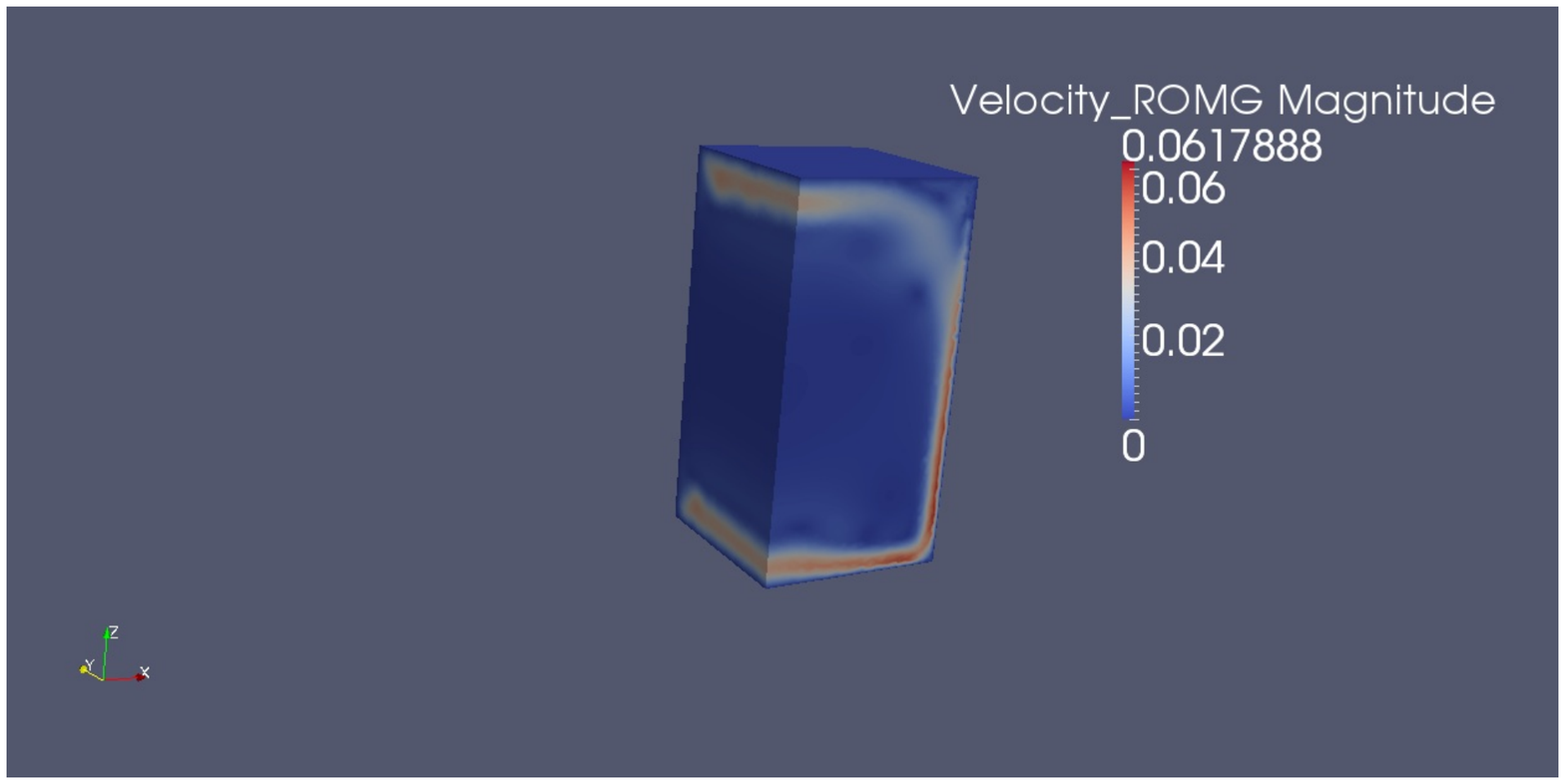}
  \vspace{-8cm}\caption{ROM-G-Learning velocity profile}
  \label{PODGLV}
\end{figure}

\begin{figure}
 \center
\includegraphics[width=0.5\linewidth]{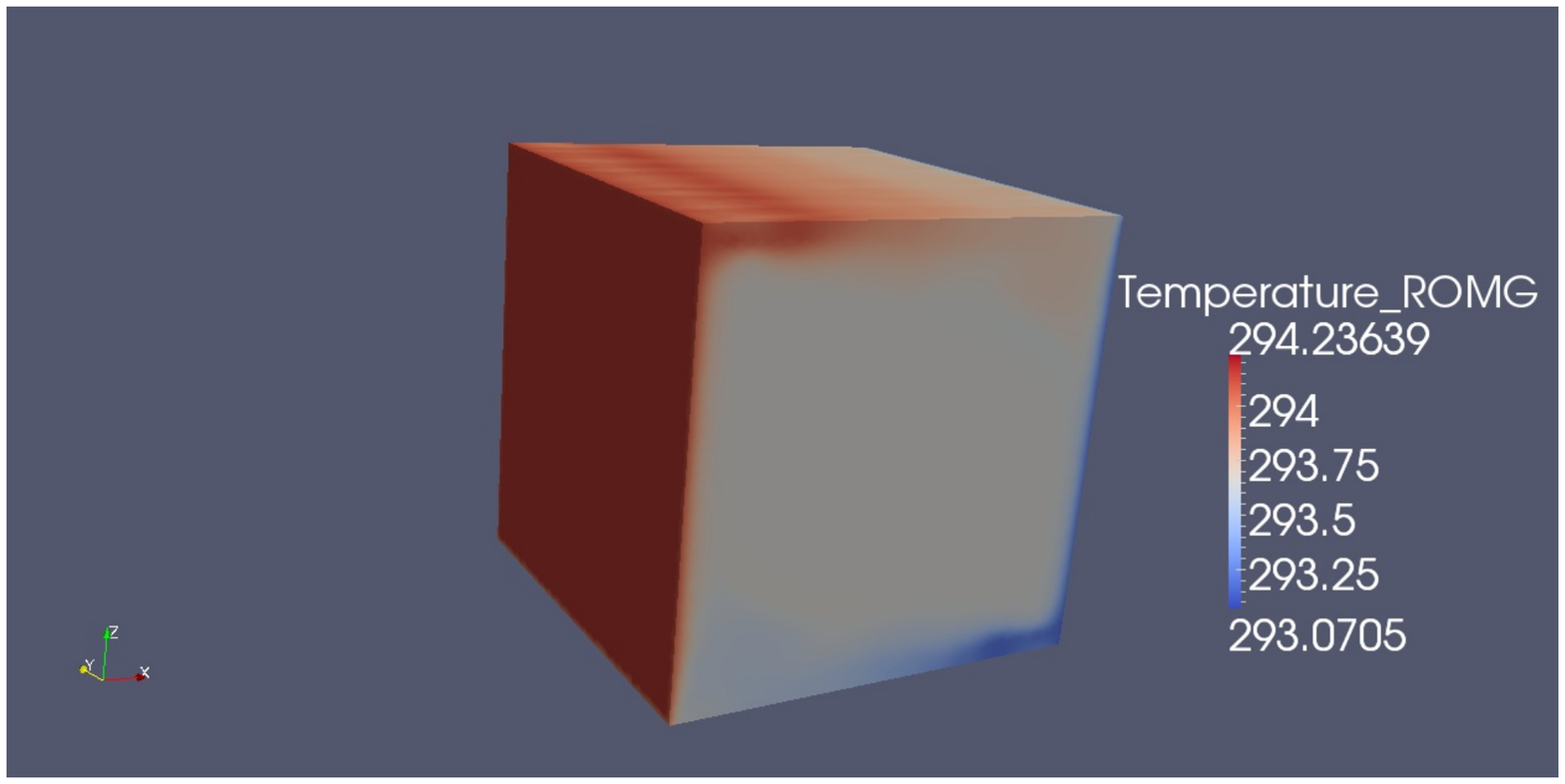}
  \vspace{-8cm}\caption{ROM-G-Learning temperature profile}
  \label{PODGLT}
\end{figure}

\begin{figure}\center
\includegraphics[width=0.5\linewidth]{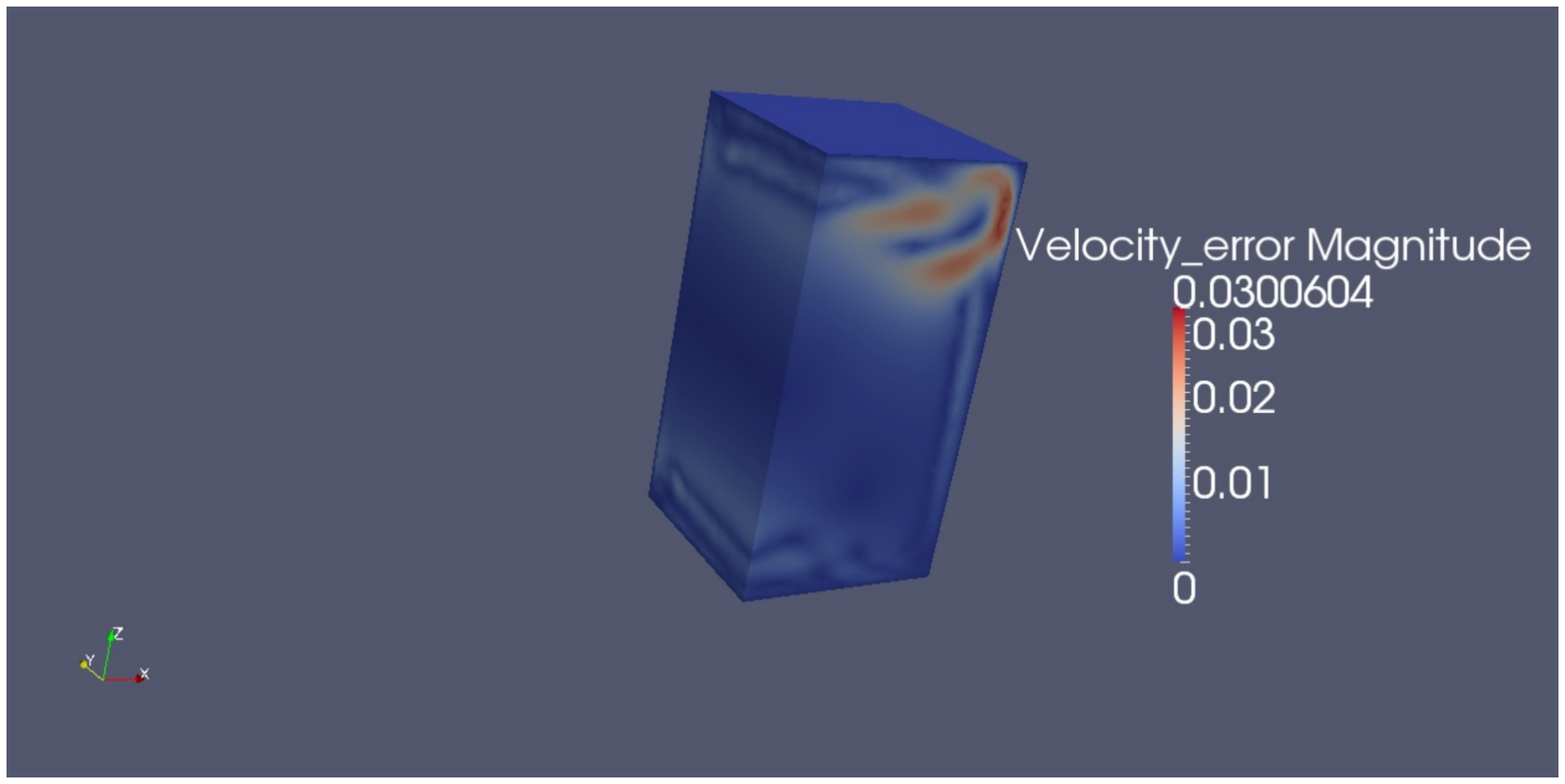}
  \vspace{-8cm}\caption{ROM-G-Learning velocity error profile}
  \label{errorVGL}
\end{figure}

\begin{figure}\center
\includegraphics[width=0.4\linewidth]{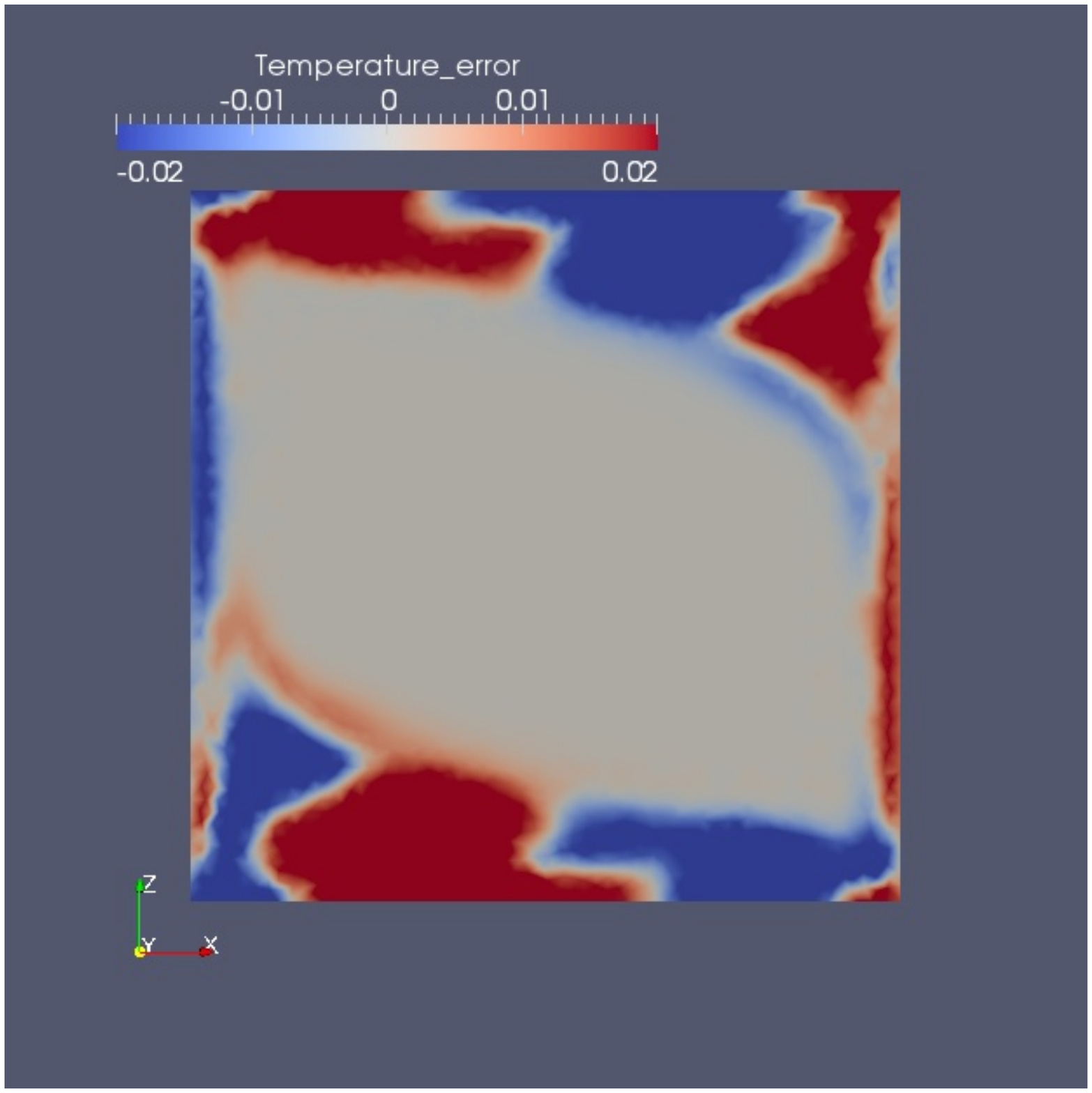}
  \vspace{-3cm}\caption{ROM-G-Learning temperature error profile}
  \label{errorTGL}
\end{figure}

\begin{figure}
\center \vspace{-4cm}
\hspace{-1cm}\includegraphics[width=10cm,height=14cm]{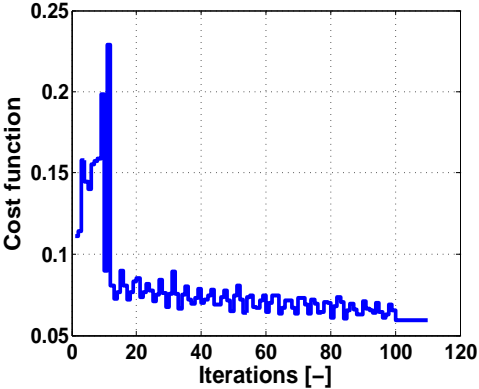}
  \vspace{-4cm}\caption{Learning cost function vs. number of learning iterations}
  \label{Q}
\end{figure}

\begin{figure}
\center\vspace{-3cm}
\hspace{-1cm}\includegraphics[width=10cm,height=14cm]{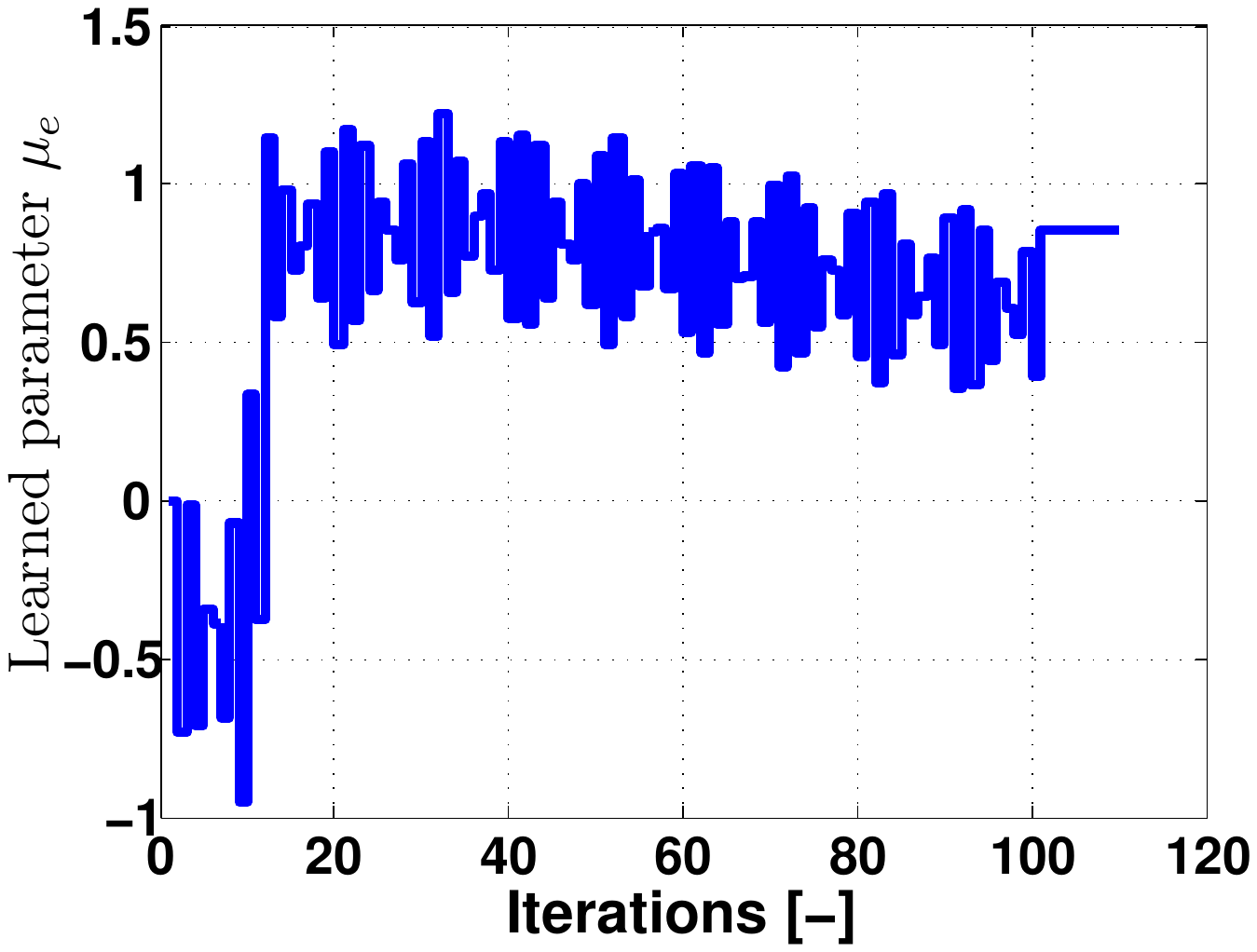}
  \vspace{-4cm}\caption{Coefficient $\mu_e$ vs. number
of learning iterations}
  \label{hatnue}
\end{figure}

\begin{figure}
\center \vspace{-4cm}
\hspace{-1cm}\includegraphics[width=10cm,height=14cm]{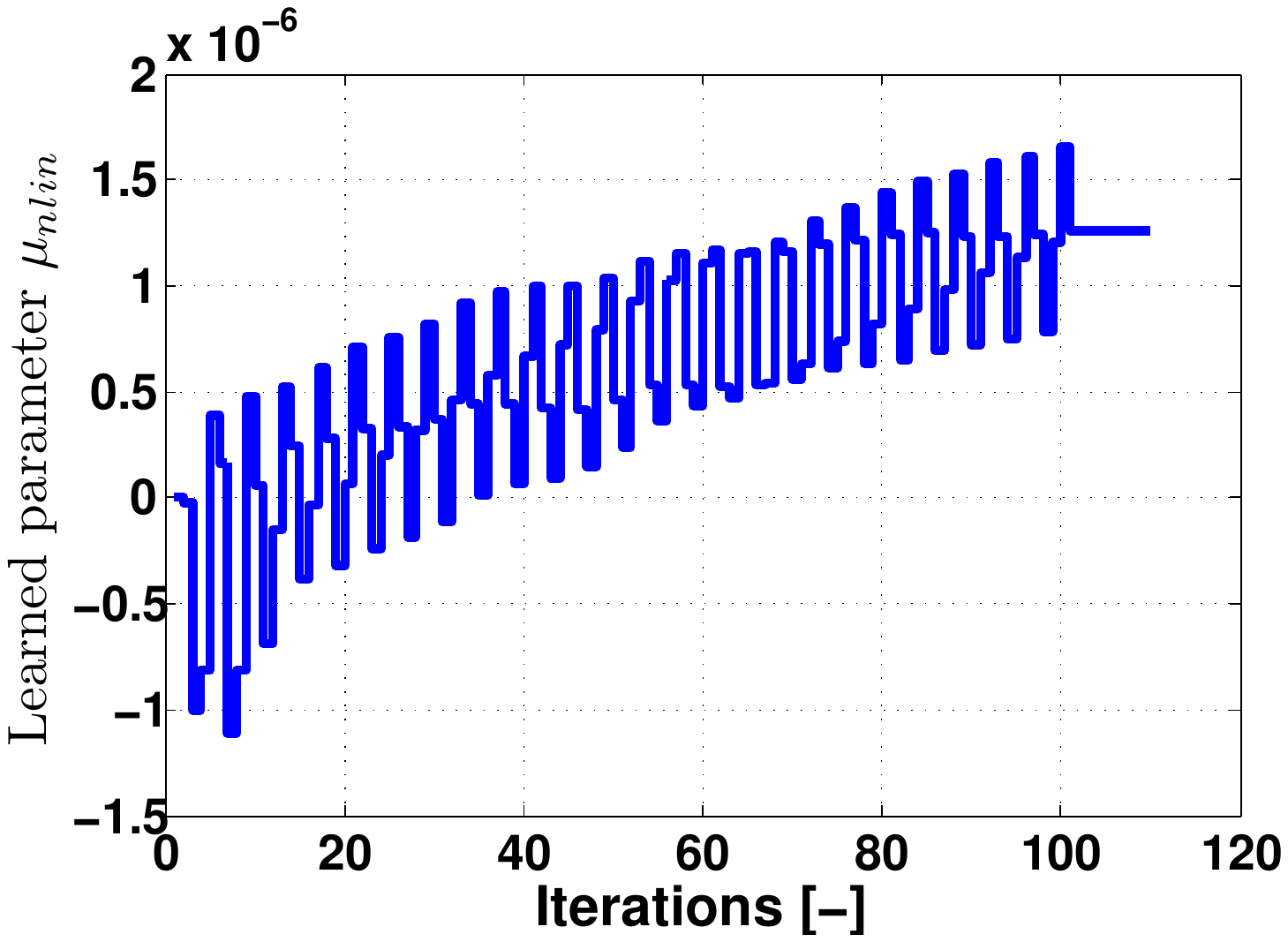}
  \vspace{-4cm}\caption{Coefficient $\mu_{nl}$ vs.
number of learning iterations}
  \label{hatnunl}
\end{figure}

Next, we show the profile of the learning cost function over the
learning iterations in Figure \ref{Q}. We can see a quick decrease
of the cost function within the first $20$ iterations. This means
that the MES manages to improve the overall solutions of the
POD-ROM very quickly. The associated profiles for the two learned
closure model amplitudes $\widehat{\mu}_{e}$ and
$\widehat{\mu}_{nl}$ are reported in Figures \ref{hatnue}, and
\ref{hatnunl}. We can see that even though the cost function value
drops quickly, the MES algorithm continues to fine-tune the values
of the parameters $\widehat{\mu}_{e}$, $\widehat{\mu}_{nl}$ as the
simulation proceeds, and eventually reach optimal values of
$\widehat{\mu}_{e}\simeq 0.85$, and $\widehat{\mu}_{nl}\simeq
1.25e-6$ when convergence tolerances are met. We also show the
effect of the learning on the POD-ROM solutions by plotting the
errors $e_T$ and $e_{\bf v}$ in Figures \ref{PODGLV},
\ref{PODGLT}, \ref{errorVGL}, \ref{errorTGL}, which by comparison
with Figure \ref{PODGV}, \ref{PODGT}, \ref{errorVG}, \ref{errorTG}
show an improvement of the POD-ROM solutions with the MES tuning
of the closure models' amplitudes.


\newpage
\section{Conclusion}\label{concl}
In this work we have proposed a new closure model for ROMs that
provide robust stabilization when applied to PDEs with parametric
uncertainties. We have also proposed the use of a model-free
multi-parametric extremum seeking (MES) algorithm to auto-tune the
closure model coefficients that optimize the POD-ROM solution
predictions. We have validated the proposed method on a
challenging 3D Boussinesq test-case by considering a
Rayleigh-B\'enard differentially-heated cavity problem. The
proposed closure model has shown encouraging performance in terms
of improving solution precision in the laminar flow cases
considered here. Future investigations will be conducted on more
challenging flows, e.g., turbulent flows, and online experimental
tests using a water-tank test-bed.

\end{document}